\documentclass[letterpaper,12pt]{article}

\usepackage[latin1]{inputenc}
\usepackage{url}

\advance\oddsidemargin by -10mm
\advance\textwidth by 10mm

\title{Steady $f$-plane circulation in basins with saddle-point
bathymetry\\[5mm]
{\normalsize Submitted to {\em Journal of Physical Oceanography},
2005-01-12.\\
Corrected 2005-04-15.}
}
\author{Alastair D.~Jenkins\thanks{%
Bjerknes Centre for Climate Research,
Geophysical Institute, Allégaten 70, 5007 Bergen, Norway.  E-mail:
alastair.jenkins@bjerknes.uib.no
}}
\date{}
\begin{document}
\maketitle
\begin{abstract}
Nilsson {\em et al.}\ have recently shown how a velocity field in geostrophic
and hydrostatic balance in an $f$-plane may be diagnosed from prescribed
distributions of buoyancy and wind stress, in a basin with closed isobaths.
I extend their analysis to cover
basins with more complex depth contours, treating in particular the behavior of
the flow in the presence of a saddle point.  
\end{abstract}
\section{Introduction}
Recently, Nilsson {\em et al.}~\cite{NilssonJ-WalinG-BrostromG:jmr-2005-xxx} showed that the
steady circulation in a  basin with closed depth 
contours, under conditions of geostrophic
and hydrostatic balance, with arbitrarily prescribed distributions of buoyancy
and wind stress, may be determined by integral constraints, obtained by
integrating around the contours.  Taking account of bottom friction, 
they found that
the net Ekman transport across the closed contours must vanish, and that this
constraint determines the free barotropic motion and the entire velocity field.
If, however, the depth contours are not all simple closed curves, but include
one or more saddle points, the integral constraints may require the fluid
velocity to be discontinuous on contours which include saddle points.  This
discontinuity they showed to be inconsistent with their scaling assumption that 
the flow and the bathymetry vary over comparable length scales.

In this paper we consider the steady flow in a basin which contains a 
saddle point.  In the presence of linear bottom friction, the governing 
partial differential equation for the barotropic motion is 
shown to be elliptic and non-singular if the depth topography is smooth and has
a bounded gradient. 
If we investigate the behavior of the solutions to the equation in the
vicinity of a saddle point 
\[
H(x,y) = H_0 + Axy,
\]
we find that the equation is separable, and the solution may be expressed as a
sum or integral of solutions to Hermite's equation.

\section{Mathematical development}

For a basin with variable depth $H$ and with surface and bottom Ekman layers,
containing water of variable density
$\rho_r(1-q)$, Nilsson {\em et al.}~\cite{NilssonJ-WalinG-BrostromG:jmr-2005-xxx} derived the
following equation for the steady, geostrophic current ${\bf u}$ in the
interior:
\begin{equation}
{\bf u}(x,y,z) = {g\over f}{\bf R}\int_{-H}^z\nabla_h q\,dz + {g\over f}q_b{\bf
R}\nabla H + {1\over f\rho_r}{\bf R}\nabla p_0,
\label{eq-u}
\end{equation}
where $g$ is the acceleration due to gravity, $f$ is the Coriolis parameter,
$q_b=q(x,y,-H)$, $p_0$ is a barotropic pressure anomaly, and the linear
operator ${\bf R}$ rotates vectors 90~degrees to the left in the $xy$-plane. 
Nondimensionalizing so that $\rho_r=g=f=1$, we obtain, for the bottom velocity
${\bf u_b}$ (just outside the Ekman layer):
\begin{equation}
{\bf u_b} = q_b{\bf R}\nabla H + {\bf R}\nabla p_0.
\label{eq-ub}
\end{equation}

Nilsson {\em et al.}~\cite{NilssonJ-WalinG-BrostromG:jmr-2005-xxx} also show (using our
non-dimensional notation) that
\begin{equation}
{\bf u_0}\cdot\nabla H = \nabla\cdot({{\bf u_0} H}) = -\nabla\cdot{\bf m},
\label{eq-u0-m}
\end{equation}
where
\begin{equation}
{\bf u_0} = {\bf R}\nabla p_0,
\end{equation}
and the Ekman boundary-layer transport ${\bf m}$ is given by
\begin{equation}
{\bf m} = -{\bf R}(\tau_w-\tau_b),
\end{equation}
where ${\tau_w}$ is the wind stress and $\tau_b$ is the bottom stress, given,
in a linear representation, by
\begin{equation}
\tau_b = h_e{\bf u_b},
\label{eq-taub}
\end{equation}
where $h_e$ is the Ekman depth.

From (\ref{eq-u}--\ref{eq-taub}), we obtain
\[
\nabla \cdot (H{\bf R}\nabla p_0) = \nabla\cdot\left[{\bf R}\tau_w -
h_e{\bf R}\left(q_b {\bf R} \nabla H + {\bf R}\nabla p_0 \right)
\right]
,
\]
which we may write as
\begin{equation}
\nabla\cdot\left[(H{\bf R}-h_e{\bf I})\nabla p_0\right]
= \nabla\cdot\left(h_e q_b\nabla H\right) - \nabla\times\tau_w
,
\label{eq-diveq}
\end{equation}
${\bf I}$ being the identity operator, or
\begin{equation}
h_e\nabla^2 p_0 + \nabla H \times \nabla p_0 = \nabla \times \tau_w -
h_e\left(q_b\nabla^2 H + \nabla q_b \cdot \nabla H \right).
\label{eq-p0}
\end{equation}
For non-zero $h_e$, (\ref{eq-p0}) is elliptic.

In the absence of friction ($h_e = 0$), $\nabla H \times \nabla p_0 = 0$, and
$p_0$ may be an arbitrary, not necessarily continuous, function of~$H$ if the
forcing terms on the right-hand side of (\ref{eq-p0}) are zero.  The flow
velocity is directed along the depth contours.

For $0<h_e\ll H$, if the depth contours are closed,
Nilsson {\em et al.}~\cite{NilssonJ-WalinG-BrostromG:jmr-2005-xxx} showed that ${\bf u_0}$ is
given by
\begin{equation}
{\bf u_0} = -\left[{\hat q}(H)+{\hat T}(H)\right]{\bf R}\nabla H,
\label{eq-nilsson-u0}
\end{equation}
where 
\begin{equation}
{\hat q} = \oint_{C(H)} q_b \left|\nabla H\right| ds \left(\oint_{C(H)}
\left|\nabla H\right| ds\right)^{-1}
\end{equation}
and
\begin{equation}
{\hat T}(H) = {h_e}^{-1}\oint_{C(H)}\tau_w\cdot d{\bf s}
\left(\oint_{C(H)} \left|\nabla H\right| ds\right)^{-1}
,
\label{eq-That}
\end{equation}
where integration is around the depth contour $C(H)$.  If the bottom contours
are all smooth closed curves, with no saddle points, and $\tau_w$ is smooth,
then $p_0$ and the flow velocity will be smooth and continuous.  If, however,
the bottom profile has one or more saddle point, it is possible for
(\ref{eq-nilsson-u0}--\ref{eq-That}) to give flow velocities which are
discontinuous across a separatrix (the isobaths which touch at a saddle
 point).  This
violates the scaling assumption of 
Nilsson {\em et al.}~\cite{NilssonJ-WalinG-BrostromG:jmr-2005-xxx} that the flow and bathymetry
vary over comparable length scales, and it is likely that the solution to
(\ref{eq-p0}) will have a boundary-layer structure in the vicinity of the
separatrix.

Note that the zero bottom slope at a saddle point should not lead directly 
to any
singularities in the solution, since
if $\nabla H = 0$, (\ref{eq-p0}) reduces to the Poisson equation
\begin{equation}
h_e\nabla^2 p_0 = \nabla\times\tau_w.
\end{equation}

\section{Solution in the vicinity of a saddle point}
To determine the behavior of solutions to~(\ref{eq-p0}) in the presence of a
saddle point, we consider a special case of the homogeneous version
of~(\ref{eq-p0}), with $H = H_0 + Axy$, 
the saddle point being at the origin. We drop
the zero suffix from $p_0$ for convenience.  
We have
\begin{equation}
{h_e}\nabla^2 p - Ax{\partial p\over\partial x} 
+ Ay {\partial p\over\partial y} = 0.
\label{eq-saddle}
\end{equation}
This linear partial differential
equation is separable, with solutions of the form $X(x) Y(y)$, satisfying
versions of the Hermite equation
\begin{equation}
{h_e\over A}X'' - xX' + kX = 0;\qquad {h_e\over A}Y'' + yY' - kY = 0.
\label{eq-saddle-XY}
\end{equation}
There is thus the possibility of constructing boundary layer solutions by
employing linear combinations of solutions to~(\ref{eq-saddle-XY}) for suitable
combinations of values of~$k$.

\section{Boundary-layer structure}
\subsection*{\it a. inner solution}
If we let $x^* = x/L$ and $y^* = y/L$ in~(\ref{eq-saddle}--\ref{eq-saddle-XY}), where 
$L=(2{h_e/A})^{1/2}$, we obtain,
removing the asterisks for convenience:
\begin{equation}
\nabla^2p - 2x{\partial p\over\partial x} + 2y {\partial p\over\partial y}
= 0,
\label{eqscaled-p}
\end{equation}
and
\begin{eqnarray}
{d^2X\over {dx}^2} - 2x{dX\over dx} + 2kX &=& 0;\label{eq-h1}\\
{d^2Y\over {dy}^2} + 2y{dY\over dy} - 2kY &=& 0.
\label{eq-hermite}
\end{eqnarray}
Equations~\ref{eq-h1}--\ref{eq-hermite} are variants of Hermite's equation, 
whose solutions for
integer values of~$k$ include Hermite polynomials and repeated integrals of the
complementary error function~\cite{AbramowitzM-StegunI:HMF-1965}.

For $k\geq0$, Eq.~\ref{eq-hermite} has solutions of the form 
\begin{eqnarray}
X_k &=& A_1 H_k(x) + B_1{\cal I}^{k+1} e^{{x}^2},\nonumber\\
Y_k &=& A_2{\cal I}^k{\rm\,erfc}(y) + B_2{\cal I}^k{\rm\,erfc}(-y),
\end{eqnarray}
where $A_1$, $B_1$, $A_2$, and $B_2$ are constants, $H_k$ is the Hermite
function of order $k$ (reducing to the Hermite polynomial of degree $k$ if $k$
is an integer),
\begin{equation}
{\rm\,erfc}(z) = {2\over\sqrt{\pi}}\int_z^\infty e^{-z^2}
\end{equation}
is the complementary error function, and ${\cal I}^k$
denotes the following integral operator applied $k$
times:
\begin{equation}
{\cal I}f(x) = \int_x^\infty f(t)\,dt,
\end{equation}
suitably generalized for non-integer $k$. 
For $k<0$, the solutions are
\begin{eqnarray}
X_k &=& C_1{\cal I}^{-k}{\rm\,erfc}(-x) + D_1{\cal
I}^{-k}{\rm\,erfc}(x),\nonumber\\
Y_k &=& C_2 H_{-k}(-y) + D_2{\cal I}^{-k+1} e^{{y}^2},
\end{eqnarray}
where $C_1$, $D_1$, $C_2$, and $D_2$ are constants.

For the particular cases $k=\pm1$ with $B_1=D_2=0$, and $A_1=C_2=1$, 
we obtain
\begin{eqnarray}
X_1 &=& x, \nonumber \\
Y_1 &=& A_2 \left(-y {\rm\,erfc}(y) + (\pi)^{-1/2}e^{-y^2}\right)
+ B_2\left(y {\rm\,erfc}(-y) - (\pi)^{-1/2}e^{-y^2}\right);
\nonumber\\
X_{-1} &=& 
C_1 \left(x {\rm\,erfc}(-x) - (\pi)^{-1/2}e^{-x^2}\right)
+ D_1 \left(-x {\rm\,erfc}(x) + (\pi)^{-1/2}e^{-x^2}\right),\nonumber\\
Y_{-1} &=& y;
\label{eq-inner}
\end{eqnarray}
where we have employed the identities 
$H_{1}(z) = z$ and ${\cal I} {\rm\,erfc}(z) = - z {\rm\,erfc}(z) + (\pi)^{-1/2}e^{-z^2}$.

\subsection*{\it b. matching inner and outer solutions}
The inner solution 
\begin{equation}
p^{i}(x,y) =
X_1(x,y)Y_1(x,y)+X_{-1}(x,y)Y_{-1}(x,y),
\label{inner-solu}
\end{equation} in the
limit of large $|x|$ and $|y|$, tends to the following functional form:
\begin{eqnarray}
\!\!\!\!\hbox{For $y>0$:}&&\!\!\! 2(B_2-D_1)xy\quad\hbox{if}\quad x<0,\quad
2(B_2+C_1)xy\quad\hbox{if}\quad x>0;\nonumber \\
\!\!\!\!\hbox{For $y<0$:}&&\!\!\! 2(-A_2-D_1)xy\quad\hbox{if}\quad x<0,\quad
2(-A_2+C_1)xy\quad\hbox{if}\quad x>0.
\label{outer-limit}
\end{eqnarray}

We see that this outer limit of $p^i$
is proportional to $xy$ in each of the
four quadrants,
The coefficient of proportionality may be
specified
independently in three of the four
quadrants by choosing a 
suitable combination of
$A_2$,
$B_2$, $C_1$, and $D_1$.  The fluid
velocity at the outer scale
may have two independently-specified
discontinuities along the positive
and negative $x$ and $y$ axes.  The third
independent
discontinuity depends on the other two
(the sum of all the
discontinuities must of course equal
zero).  Since the integral
condition
(\ref{eq-nilsson-u0}-\ref{eq-That}) may
specify four
independent proportionality conditions,
and thus three independent
velocity discontinuities, we do not yet
have a complete agreement
between the inner and outer solutions.

\section{Conclusion}
The fact that one may have, in the outer
limit, three
 independently-specified
discontinuities along the positive
and negative $x$ and $y$ axes in the
vicinity of the bathymetric saddle point,
means that it is possible to
specify different integral constraints
for the circulation in  
three of
the four quadrants.  Extending this
result to the larger-scale flow
pattern in a basin, this means that flow
velocity discontinuities predicted from
(\ref{eq-nilsson-u0}--\ref{eq-That})
across separatrices in the isobaths
are not quite resolvable at an inner
scale by solutions of the form given in
(\ref{eq-inner}--\ref{inner-solu}) in
terms of the current
analysis.  Further investigation is
required in order to obtain
complete agreement between the solutions
at the inner and outer
scales.

The solutions calculated in this paper
are valid in the vicinity of
saddle points which have separatrices
crossing at right angles.  For
more general types of saddle point, the
homogeneous version
of~(\ref{eq-p0}) is no longer
separable: however, I anticipate that the
matching of inner and outer
solutions will eventually allow for the
appropriate specification of
outer-scale flow discontinuities across
separatrices. 
Hence the
integral-constraint specification of the steady circulation proposed
by Nilsson {\em et al.}~\cite{NilssonJ-WalinG-BrostromG:jmr-2005-xxx} should be valid even
for bathymetries containing saddle points.

\section*{Acknowledgments}
I thank Johan Nilsson for valuable discussions and for providing me with an
unpublished version of the Nilsson {\em et al.}~\cite{NilssonJ-WalinG-BrostromG:jmr-2005-xxx}
manuscript.
The work was supported by the Research Council of Norway
under project no.~155923/700.
This is publication no.~0000 of the Bjerknes Centre for Climate
Research.

\bibliographystyle{habbrv}

\end{document}